\documentstyle[aps,prl,multicol,epsf]{revtex} 
\begin{document} 
\widetext
 
\title{Non-linear feedback effects in coupled Boson-Fermion systems}

\author{T.~ Domanski$^{(a,b)}$ and J.~Ranninger$^{(a)}$}
\address{$^{(a)}$ Centre de
Recherches sur les Tr\`es Basses Temp\'eratures, Laboratoire
Associ\'e \'a l'Universit\'e Joseph Fourier, Centre National de la
Recherche Scientifique,\\ BP 166, 38042, Grenoble C\'edex 9, France}
\address{$^{(b)}$ Institute of Physics, Marie Curie Sklodowska University, 
20031 Lublin, Poland.}
\date{\today}  
\maketitle 
\draft 
\begin{abstract}
We address ourselves to a class of systems composed of two coupled 
subsystems without any intra-subsystem interaction: itinerant Fermions
 and localized Bosons on a lattice.  
Switching on an interaction between the two subsystems leads to feedback 
effects which result in  a rich dynamical structure in both of them. Such 
feedback features are studied on the basis of the flow equation 
technique - an 
infinite series of infinitesimal unitary transformations - which leads 
to a gradual elimination of the inter-subsystem interaction.
As a result the two subsystems get decoupled but their 
renormalized  kinetic energies become mutually dependent on each 
other. Choosing for the inter- subsystem interaction a charge exchange 
term, - the $\it{Boson-Fermion \; model}$
- the initially localized Bosons acquire itinerancy through their 
dependence on the renormalized Fermion dispersion. This latter evolves 
from a free
particle dispersion into one showing a pseudogap structure near the 
chemical potential. Upon lowering the temperature both subsystems 
simultaneously enter a macroscopic coherent quantum state. The Bosons 
become superfluid, exhibiting a soundwave like dispersion while the 
Fermions develop a true gap in their dispersion. The essential physical 
features described by this technique are already contained in the 
renormalization of the kinetic terms in the respective Hamiltonians of 
the two subsystems. The extra interaction terms resulting in the process 
of iteration only  strengthen this physics. We compare the results 
with previous calculations based on selfconsistent  perturbative 
approaches.  
\end{abstract}

\begin{multicols}{2}

\narrowtext
\section{Introduction}
A wide class of problems in solid state physics can be described 
 in terms of interacting Boson - Fermion systems. Examples 
are:

i) the electron-phonon problem and its associated realization in 
superconductivity and polaron formation,

ii) fermions interacting with spin fluctuations, relevant for the 
description of heavy fermion systems as well as for high temperature 
superconductors,

iii) certain systems which can be mapped into  Boson-Fermion interacting 
systems via Hubbard Stratanovich transformations,

iv) the Anderson impurity and Kondo problem and finally

v) the  Boson-Fermion model (BFM) for high $T_c$ superconductivity,
believed to describe a coupled electron-phonon system in 
the cross-over regime between weak and strong coupling.

In order to obtain the correct low energy physics in these various scenarios 
of Boson-Fermion interacting systems the mutual feedback effects caused 
by the interaction between the two subsystems must be handled properly. 
It invariably gives rise to  effective time dependent interactions among 
the constituents in each subsystem which quite  generally can be obtained 
by the standard field theoretical method based on functional integrals. 
This method is particularly suited 
if one considers the  limit of infinite dimensions where it has been 
developed in great detail and is known as dynamical mean field 
theory\cite{George-96}. The mutual 
feedback effects have been recently studied on the basis of this 
method for the Boson-Fermion model\cite{Robin-98} and for the many
polaron problem in the Holstein model\cite{Kotliar-00}.
While this method is non-perturbative and 
capable of describing the low energy physics,  it is, as 
up to now, restricted to the study of local quantities.
In situations where the dimensionality and anisotropy of the 
physical systems play a role, such as believed to be the case in 
the high $T_c$ cuprates, one has to resort to different techniques 
to handle these feed back effects. There are many physical situations 
in which already a relatively small 
inter-subsystem coupling leads to substantial fluctuations in each of 
them and which hence requires the selfconsistent determination of 
the mutual non-linear feedback effects between the Fermions and 
Bosons. In those cases perturbative approaches in form of 
selfconsistent diagrammatic techniques can and have been 
applied for the electron-phonon problem  
(the Eliashberg approach\cite{Eliashberg-60}) and 
for the BFM\cite{Ranninger-95}. These approaches  however
totally neglect vertex corrections.

Controlled perturbative methods in the spirit of renormalization 
group techniques have been recently proposed, the 
$\it{similarity \; renormalization \; scheme}$\cite{Glazek-94} and the 
$\it{flow \; equation \; technique}$\cite{Wegner-94}. These techniques 
 are capable in reformulating such interacting systems in terms 
of renormalized Hamiltonians which capture the low energy physics, which
 is achieved via an infinite series of infinitesimal unitary 
transformations. Contrary to the standard simple unitary 
transformations which treat the different energy scales in the 
problem in a single step (and therefore generally fail) these continuous 
transformations deal with each energy scale in a sequence of 
transformations and by doing so are capable of extracting the 
low energy physics in  which we are interested.

This paper is organized in the following way. In the subsequent 
section II we review the essential points of the flow equation technique 
and apply it to the Boson-Fermion model. In section III we discuss 
the excitation spectrum of the renormalized Hamiltonian and compare 
the results to those previously obtained by different methods. In 
section IV we give a preliminary discussion of the superconducting 
phase properties of the BFM.

Apart from testing this flow equation method for this model, this present 
study on the BFM goes beyond the studies so far reported in the past. For 
the first time we are able to treat on the same footing the normal and the 
superconducting phase and  handle the question how the pseudogap evolves 
into a true gap below $T_c$.

 \section{The flow equation technique}

The method of {\em continuous} unitary transformation has been
formulated by G\l azek and Wilson \cite{Glazek-94} and independently
by Wegner \cite{Wegner-94}. Instead of the single-step transformation
this method amounts to a procedure for simplifying the Hamiltonian's 
representation through the series of infinitesimal unitary
transformations $U(l)$, where $l$ denotes the continuous {\em  flow parameter}.

The continuous unitary transformation $H(l)=U^{\dagger}(l)HU(l)$ 
gives rise to the following {\em flow equation} 

\begin{equation}
\frac{dH(l)}{dl} = \left[ \eta(l),H(l) \right] 
\label{flow}
\end{equation}
where $\eta(l)$ represents some  arbitrary (anti-hermitean) generator
\begin{equation}
\eta(l) = \frac{dU^{\dagger}(l)}{dl}U(l) \;.
\end{equation}

The choice of a specific form of $\eta(l)$ is usually
determined by the physical situation under consideration
\cite{Glazek-94,Wegner-94,Mielke-98}. In this paper we use
Wegner's proposal \cite{Wegner-94} 
\begin{equation}
\eta(l) = \left[ H_{0}(l), H_{int}(l) \right]
\label{wegner}
\end{equation}
where $H_{0}$ is the diagonal and $H_{int}$ the nondiagonal part of 
the Hamiltonian in a given representation. In general $H_{int}$ will
be understood as the perturbation with respect to $H_{0}$.
The generating operator Eq. (3) guarantees that under
the continuous transformation the off diagonal terms are monotonously
reduced, eventually leading to a block diagonalization of the Hamiltonian,
provided that no degenerancies are encountered \cite{Wegner-94,Kehrein-94}. 
Recently Mielke has proposed \cite{Mielke-98} some different form of the 
$\eta(l)$ operator which can be used for studying systems with degeneracies.
 
One of the first condensed matter problems analyzed with use of 
the flow equation was the electron-phonon Hamiltonian\cite{el-ph}. 
It has been shown that eliminating the electron-phonon interaction
induces the effective interactions between electrons which are 
attractive in the whole Brillouin zone. Near the Fermi surface
this attraction is strongly enhanced but it never becomes divergent
as in the case for the classical Fr\"ohlich transformation. 
The method has also been successfully applied to a variety 
of other physical problems like the single impurity Anderson
model \cite{Kehrein-96} (the Schrieffer - Wolff transformation
has been improved), the strong coupling expansion for the 
Hubbard model \cite{Stein-97} ($t/U$ expansion), the large 
spin Heisenberg Hamiltonian  \cite{Stein-98} ($1/S$ expansion),
and for other  topics  such as dissipative quantum
systems \cite{spin-boson}, light front QCD \cite{QCD},
quarkonium spectra \cite{Brisudova-97} and the Sine-Gordon 
model\cite{Kehrein-99}. 

The main advantage of the continuous transformations is that an 
 effective Hamiltonians can be derived which is valid  not only 
in the low energy sector
(like the standard Renormalization Group) but in the overall
regime of energies. In principle it is also possible to formulate
the flow equations in such a way that lifetime effects 
can  be studied too. Such type of flow equations have been 
used so far to account for dynamical effects in the context of spin-boson
problem \cite{spin-boson} and also in the investigation of phonon damping 
effects due to the electron-phonon interaction \cite{Ragawitz-99}.

\vspace{1cm}
{\bf A. Application to the Boson-Fermion model}

\vspace{5mm}
 We shall now apply this flow equation technique to the BFM
described by the following Hamiltonian 
\begin{equation}
H = H_{0} + H_{int} .
\label{BF}
\end{equation}
The free part (diagonal in the basis of the plane waves)
consists of  the kinetic terms of the Fermions and Bosons
\begin{eqnarray}
H_{0} = \sum_{k,\sigma} (\varepsilon_{k}^{\sigma} - \mu)
        c_{k\sigma}^{\dagger}c_{k\sigma} +
        \sum_{q} (E_{q}-2\mu) b_{q}^{\dagger}b_{q} .
\end{eqnarray}
The mutual interaction between both species is represented by 
the charge exchange term 
\begin{eqnarray}
H_{int} = \frac{1}{\sqrt N} \sum_{k,p} \left( v_{k,p} b^{\dagger}_{k+p} 
c_{k\downarrow}c_{p\uparrow} + v^{*}_{k,p} b_{k+p}
c^{\dagger}_{p\uparrow}c^{\dagger}_{k\downarrow} \right).
\end{eqnarray}
The flow equations control the evolution of the  model parameters 
$\varepsilon^{\sigma}_{k}$,
$E_{q}$, $v_{k,q}$ which  get renormalized in the course of 
the flow equations procedure. 
>From now on we assume that they depend on
the flow parameter $l$ with the following initial conditions
\begin{eqnarray}
E_{q}(l=0) & = & \Delta_{B} \;,  
\varepsilon^{\sigma}_{k}(l=0) = \epsilon_{k} \;, 
v_{k,p}(l=0)=v \;.
\label{initial_conditions}
\end{eqnarray}
According to  Wegner's definition (\ref{wegner}) of
the generating operator we have
\begin{eqnarray}
\eta(l) & = &  \frac{1}{\sqrt N} \sum_{k,p} 
\left( \varepsilon^{\downarrow}_{k}(l) +
\varepsilon^{\uparrow}_{p}(l) - E_{k+p}(l) \right)\nonumber \\
&\times& \left( v^{*}_{k,p}(l) b_{k+p} c_{p\uparrow}^{\dagger}
c_{k\downarrow}^{\dagger} - v_{k,p}(l) b_{k+p}^{\dagger} 
c_{k\downarrow}  c_{p\uparrow} \right)
\label{generator}
\end{eqnarray}

Upon iterating  the flow equations (\ref{flow})  new interaction terms
 are in general created which are not contained in the original Hamiltonian. 
Certain of those terms can be eliminated  by  reformulating the initial 
$H_{0}$ in the following way 
\begin{eqnarray}
H_{0}(l) & = & \sum_{k,\sigma} \left( \varepsilon_{k}^{\sigma}(l)
- \mu \right)  : c_{k\sigma}^{\dagger}c_{k\sigma} :  +
\sum_{q} \left( E_{q}(l)-2\mu \right) :b_{q}^{\dagger}b_{q}:
\nonumber \\ & + &
\frac{1}{N}\sum_{p,k,q} U_{p,k,q}(l) c^{\dagger}_{p\uparrow}
c^{\dagger}_{k\downarrow} c_{q\downarrow} 
c_{p+k-q\uparrow} + {\rm c}(l)
\label{renorm}
\end{eqnarray}
where the symbol $:x:=x-<x>$ stands for the normal order product 
and ${\rm c(l)}$ denotes a c-number contribution to the Hamiltonian.
 We furthermore 
supplement the initial conditions (\ref{initial_conditions}) with  the
constraints
\begin{eqnarray}
&U&_{p,k,q}(l=0)  =  0 \;,
\label{constraints1} \\
&{\rm c}&(l=0)  =  \sum_{k,\sigma} \left( \varepsilon_{k} - \mu \right)
n^{(FD)}_{\sigma,k} + \sum_{q} \left( E_{q} - 2\mu \right) n^{(BE)}_q \;.
\label{constraints2}
\end{eqnarray}
After some straight forward algebraic manipulations we finally obtain  
\end{multicols}

\widetext
--------------------------------------------------------------------------
\begin{eqnarray}
\frac{dH(l)}{dl} & = & - \frac{1}{\sqrt N} \sum_{k,p} \alpha^{2}_{k,p}(l)
\left( v_{k,p}(l)b^{\dagger}_{k+p}c_{k\downarrow}c_{p\uparrow} +
v^{*}_{k,p}(l)b_{k+p}c^{\dagger}_{p\uparrow}c^{\dagger}_{k\downarrow} \right)
\nonumber \\ & + &\frac{1}{N}
\sum_{k,p,q} \left( \alpha_{k,p}(l) + \alpha_{q,k+p-q}(l) \right)
v^{*}_{k,p}(l)v_{q,k+p-q}(l)c^{\dagger}_{p\uparrow} c^{\dagger}_{k\downarrow}
c_{q\downarrow} c_{p+k-q\uparrow}
\nonumber \\ & + &\frac{2}{N}
 \sum_{k,p} \alpha_{k,p}(l) |v_{k,p}(l)|^{2} n^{(BE)}_{p+k}
: c^{\dagger}_{k\downarrow} c_{k\downarrow} : \; + \;
\frac{2}{N} \sum_{k,p} \alpha_{p,k}(l) |v_{p,k}(l)|^{2} n^{(BE)}_{p+k}
: c^{\dagger}_{k\uparrow} c_{k\uparrow} :
\nonumber \\ & + &\frac{2}{N}
 \sum_{k,p} \left[ \alpha_{k,p}(l) |v_{k,p}(l)|^{2} 
\left( -1 + n^{(FD)}_{\downarrow ,k} \right)
+ \alpha_{p,k}(l) |v_{p,k}(l)|^{2} n^{(FD)}_{\uparrow ,k} \right]
: b^{\dagger}_{k+p} b_{k+p} :
\nonumber \\ & + &\frac{2}{N}
 \sum_{k,p} \left[ \alpha_{k,p}(l) |v_{k,p}(l)|^{2} 
\left( -1 + n^{\downarrow}_{FD}(k) \right)
+ \alpha_{p,k}(l) |v_{p,k}(l)|^{2} n^{\uparrow}_{FD}(k) \right]
n^{(BE)}_{k+p}
\nonumber \\ & + &\frac{1}{N}
\sum_{k,p,q \neq k} b^{\dagger}_{p+q}b_{p+k} \left[ \left( \alpha_{k,p}(l)+
\alpha_{q,p}(l)\right) v^{*}_{k,p}(l)v_{q,p}(l) c^{\dagger}_{k\downarrow}
c_{q\downarrow} + 
\right. \nonumber \\ & &  \left.
\left( \alpha_{p,k}(l)+\alpha_{p,q}(l) \right)
v^{*}_{p,k}(l)v_{p,q}(l) c^{\dagger}_{k\uparrow}c_{q\uparrow} \right] 
+ O(:c^{\dagger}_{k\sigma} c_{k\sigma}: :b^{\dagger}_{p}b_{p}:) +
O(v^{3})  \;.
\label{new_flow}
\end{eqnarray}
----------------------------------------------------------------------------
\begin{multicols}{2}
\narrowtext
\noindent
where for brevity we introduce
\begin{equation}
\alpha_{k,p}(l) = \varepsilon^{\downarrow}_{k}(l) +
\varepsilon^{\uparrow}_{p}(l) - E_{k+p}(l) \;.
\end{equation}

The expectation values are defined as 
\begin{eqnarray}
<c^{\dagger}_{k\sigma}c_{k\sigma}>_l & = & n^{(FD)}_{\sigma,k}(l) \equiv
\frac{1}{e^{ \left( \varepsilon^{\sigma}_{k}(l)-\mu \right) /k_{B}T}
+1} \;, \label{fermi} \\
<b^{\dagger}_{q}b_{q}>_l & = & n^{(BE)}_{q}(l) \equiv
\frac{1}{e^{ \left( E_{q}(l)-2\mu \right) /k_{B}T } -1} 
\;, \label{bose}
\end{eqnarray}
where $n^{(FD)}$, $n^{(BE)}$ are the Fermi-Dirac and Bose-Einstein
distribution functions respectively. They dependent on $l$ only through
their arguments $\varepsilon_{k}(l)$ and $E_{q}(l)$.
In order to proceed with the numerical analysis of these flow 
equations (\ref{new_flow}) we shall neglect from now on the 
terms in the last two lines. Such a neglect implies that: 

\vspace{2mm}\hspace{1cm}
\parbox{15cm}{
a)~our theory is valid up to the order $v^{2}$,\\
b)~we omit any fluctuations of the form \\
$\left( c^{\dagger}_{k\sigma}c_{k\sigma} -
\left< c^{\dagger}_{k\sigma}c_{k\sigma} \right> \right)
\left( b^{\dagger}_{p}b_{p} - \left< b^{\dagger}_{p}
b_{p} \right> \right) $,\\ 
c)~as shown in the Appendix, contributions\\
coming from the terms
$b^{\dagger}_{p+q}b_{p+k}c^{\dagger}_{k\sigma}c_{q\sigma}$ for\\
$q \neq k$ are of  order $O(v^{3})$.}

\vspace{3mm}
Within that procedure we finally arrive at the following set of the 
flow equations for the BFM
\begin{eqnarray}
\frac{dv_{k,p}(l)}{dl} & = & - \alpha^{2}_{k,p}(l) v_{k,p}(l) \;,
\label{hybrflow} \\ 
\frac{d\varepsilon^{\downarrow}_{k}(l)}{dl} 
& = & \frac{2}{N} \sum_{p} \alpha_{k,p}(l)
|v_{k,p}(l)|^{2} n^{(BE)}_{k+p}(l) \;,
\label{epsdownflow} \\
\frac{d\varepsilon^{\uparrow}_{k}(l)}{dl} 
& = & \frac{2}{N}\sum_{p} \alpha_{p,k}(l)
|v_{p,k}(l)|^{2} n^{(BE)}_{k+p}(l) \;,
\label{epsupflow} \\
\frac{dE_{k}(l)}{dl} & = & \frac{2}{N} \sum_{p}  \left[ \alpha_{k-p,p}(l) 
|v_{k-p,p}(l)|^{2} \left( -1 + n^{(FD)}_{\downarrow ,k-p}(l) \right)
\right. \nonumber \\ & + &  \left.
 \alpha_{p,k-p} |v_{p,k-p}(l)|^{2} n^{(FD)}_{\uparrow ,k-p}(l) \right] 
\label{Ekflow} \\
\frac{dU_{p,k,q}(l)}{dl} & = & \left[ \alpha_{k,p}(l) +
\alpha_{q,k+p-q}(l) \right] v^{*}_{k,p}(l)v_{q,k+p-q}(l) \;,
\label{interflow} \\
\frac{d\; {\rm c}(l)}{dl} & = & \frac{2}{N}\sum_{p} \left[ \alpha_{k,p}(l) 
|v_{k,p}(l)|^{2} \left( -1 + n^{(FD)}_{\downarrow ,k}(l) \right)
\right. \nonumber \\ & + &  \left.
\alpha_{p,k-p} |v_{p,k}(l)|^{2} n^{(FD)}_{\uparrow ,k}(l) 
n^{(BE)}_{k+p}(l) \right]\;.
\label{constflow}
\end{eqnarray}
 A formal solution for the flow equation
(\ref{hybrflow})  can be given right away
\begin{equation}
v_{k,p}(l) = v \; e^{-\int_{0}^{l} \left[ \varepsilon^{\downarrow}_{k}
(l') + \varepsilon^{\uparrow}_{p}(l') - E_{k+p}(l') \right]^{2} dl'}
\label{hybr}
\end{equation}
but of course the dispersion functions $\varepsilon^{\sigma}_{k}(l)$
and $E_{p}(l)$ ought to be determined selfconsistently via the other
flow equations.

\vspace{1cm}
{\bf B. Lowest order iterative solution}

\vspace{5mm}
The flow equation scheme is devised in such a way  that the dominant 
renormalization takes place
for the hybridization constant $v_{k,p}(l)$. To get some insight about
its effectiveness  we  solve here the flow equations
approximatively using on the right hand side of (\ref{hybrflow}-
\ref{constflow}) the bare (unrenormalized) energies $\varepsilon
^{\sigma}_{k}(l) \simeq \varepsilon_{k}$ and $E_{q}(l) \simeq 
\Delta_{B}$ (=$E_{q=0}$). 
The resulting hybridization constant (\ref{hybr}) reduces in this 
case to 
\begin{equation}
v_{k,p}(l) =  v_{p,k}(l) =  v \; e^{-\left( \varepsilon_{k} + \varepsilon_{p}
- \Delta_{B} \right)^{2} l}
\label{hybr_iter1}
\end{equation}
and  has the desired property $v_{k,p}(l \rightarrow \infty )=0$
for all  momenta, except when $\varepsilon_{k}+\varepsilon_{p}=
\Delta_{B}$. This situation corresponds to a resonant scattering 
of two Fermions into a Boson state with their total energy being 
conserved. We shall see in section III that in the selfconsistent 
solution of the flow equations such a problem does not occur. 
Substituting (\ref{hybr_iter1}) into the flow equations (\ref{epsdownflow}
-\ref{interflow})and taking the limit $l \rightarrow \infty$ one  obtains the
renormalized quantities:
\begin{eqnarray}
\varepsilon^{(R)}_{k} & = & \varepsilon_{k} + |v|^{2}\frac{1}{N}
\sum_{p} \frac{n^{(BE)}_{q=0}}{\varepsilon_{k}+\varepsilon_{p}
-\Delta_{B}} 
\label{eps_iter1}
\\
E^{(R)}_{k} & = & \Delta_{B} + |v|^{2} \frac{1}{N}\sum_{p} \frac{1 - n^{(FD)}_{p}
- n^{(FD)}_{k-p}} {\Delta_{B}-\varepsilon_{p}-\varepsilon_{k-p}}
\label{E_iter1}
\\
U^{(R)}_{p,k,q} & = & |v|^{2} \frac{\varepsilon_{k} +
\varepsilon_{p}+\varepsilon_{q}+\varepsilon_{k+p-q}-2\Delta_{B}}
{\left( \varepsilon_{k} + \varepsilon_{p} - \Delta_{B} \right)^{2} +
\left( \varepsilon_{q} + \varepsilon_{k+p-q} - \Delta_{B} \right)^{2}} 
\label{U_iter1}
\end{eqnarray}
where $\varepsilon^{(R)}_{k}$ stands for the new Fermion spectrum, 
valid for both spins.

Let us concentrate from now on on exclusively two channels 
of the induced Fermion-Fermion interactions; the zero momentum $\em BCS$ 
channel and the zero momentum density-density (d-d) channel respectively 
\begin{eqnarray}
\sum_{p,k} U^{(BCS)}_{p,k}c^{\dagger}_{p\uparrow}
c^{\dagger}_{-p\downarrow} c_{-k\downarrow}c_{k\uparrow},
\label{BCS} \\
\sum_{p,k} U^{(d-d)}_{p,k} c^{\dagger}_{p\uparrow}
c_{p\uparrow} c^{\dagger}_{k\downarrow} c_{k\downarrow}.
\label{density}
\end{eqnarray}
They denote specific elements 
$U^{(BCS)}_{p,k} \equiv U_{p,-p,-k}(l \rightarrow \infty )$
and $U^{(d-d)}_{p,k} \equiv U_{p,k,k}(l \rightarrow \infty )$
which together with Eq. (\ref{U_iter1}) become
\begin{eqnarray}
U^{(BCS)}_{p,k} & = & - |v|^{2} \; \frac{ \left( 
\Delta_{B} - 2\varepsilon_{p} \right) + \left(
\Delta_{B} - 2\varepsilon_{k} \right)  }
{\left(\Delta_{B} - 2 \varepsilon_{p} \right)^{2}+
\left(\Delta_{B} - 2 \varepsilon_{k} \right)^{2}}
\\
U^{(d-d)}_{p,k} & = & - |v|^{2} \; \frac{1}
{\Delta_{B} - \varepsilon_{p} - \varepsilon_{k} }
\end{eqnarray}
They are divergent in certain regions of the Brillouin zone, but as we
shall  show below, a selfconsistent numerical solution of the flow
equations smoothens  such divergences into a regular behavior.

\vspace{1cm}
{\bf C. Comparison with  standard perturbation the-
\indent
ory}

\vspace{5mm}
Let us next discuss the results of the first iterative solution of 
the flow equation method and compare it 
 to that obtained by  standard perturbative studies of 
the BFM, discussed in detail in Refs \cite{Ranninger-95,Ren}.
The second order expansions for the Fermion and Boson
selfenergies of the single particle propagators are 
given by
\begin{eqnarray}
\Sigma_{F}(k,i\omega_{n}) &=& \nonumber \\
|v|^{2} k_{B}&T&\frac{1}{N} 
\sum_{p,\omega_{m}} G_{F}(p-k,i\omega_{m}-i\omega_{n}) 
G_{B}(p,i\omega_{m})  \nonumber
\\
\Sigma_{B}(k,\omega_{m}) &=& \nonumber \\
 -|v|^{2} k_{B}&T&\frac{1}{N}
\sum_{p,\omega_{n}} G_{F}(k-p,i\omega_{m}-i\omega_{n})
G_{F}(p,i\omega_{n}) 
\end{eqnarray}
Our approximate solutions of the flow equations derived above 
(Section II B) were based
on unrenormalized Fermion and Boson spectra and therefore
are not selfconsistent. Let us now determine  the selfenergies 
using the bare Green's
functions, as it has been done in Ref.\ \cite{Ranninger-95} 
and which read
\begin{eqnarray}
\Sigma_{F}(k,\omega) & = & |v|^{2}\frac{1}{N} \sum_{p} 
\frac{n^{(FD)}_{p-k}+n^{(BE)}_{q=0}}
{\omega + \mu - \Delta_{B} + \varepsilon_{p-k}}, 
\label{sigmaF} \\
\Sigma_{B}(k,\omega) & = & |v|^{2}\frac{1}{N} \sum_{p} 
\frac{1-n^{(FD)}_{k-p}-n^{(FD)}_{p}}
{\omega + 2\mu - \varepsilon_{p-k} - \varepsilon_{p}}
\label{sigmaB}
\end{eqnarray}
with $n^{(BE)}_{q=0} = [ {\rm exp} \left\{ \left( 
\Delta_{B}-2\mu \right) /k_{B}T \right\} - 1 ]^{-1}$.
The effective quasi-particle spectra are then given by
the solutions of following equations
\begin{eqnarray}
\omega^{F}_{k} + \mu  - \varepsilon_{k} - \Sigma_{F}
(k,\omega^{F}_{k}) & = & 0 \;,
\\
\omega^{B}_{k} + 2\mu - \Delta_{B} - \Sigma_{B}
(k,\omega^{B}_{k}) & = & 0 \;.
\end{eqnarray}
In the limit of small $v$ we can put
\begin{eqnarray}
\omega^{F}_{k} + \mu & \simeq & \varepsilon_{k} \nonumber  \\
\omega^{B}_{k} + 2\mu & \simeq & \Delta_{B} \nonumber
\end{eqnarray}
and then substitute  these quantities into the expressions for the
 selfenergies
(\ref{sigmaF},\ref{sigmaB}) which results in
\begin{eqnarray}
\omega^{F}_{k} + \mu & \simeq & \varepsilon_{k} + |v|^{2}\frac{1}{N} 
\sum_{p} \frac{n^{(FD)}_{p-k} + n^{(BE)}_{q=0}}
{\varepsilon_{k}+\varepsilon_{p-k}-\Delta_{B}}
\label{ferm_perturb}
\\
\omega^{B}_{k} + 2\mu & \simeq & \Delta_{B}
+ |v|^{2}\frac{1}{N} \sum_{p} \frac{1-n^{(FD)}_{p}-n^{(FD)}_{k-p}}
{\Delta_{B}-\varepsilon_{p}-\varepsilon_{k-p}}
\label{bose_perturb}
\end{eqnarray}

The difference between the Fermion spectra (\ref{ferm_perturb}) 
derived from this perturbative approach and that derived from the 
flow equation approach
(\ref{eps_iter1}) can be remeded after having realized that the standard 
perturbative study describes an  effective 
Fermion spectrum while  the flow equations
method, reformulating  the Boson-Fermion interaction, results
in a) a renormalization of the Fermion energies 
$\varepsilon_{k} \rightarrow \varepsilon_{k}^{(R)}$ and
b) an appearance of the Fermion-Fermion interactions.
Taking both effects into account when
evaluating  the final Fermion quasi-particle spectrum,
 the lowest order corrections to $\varepsilon_{k}^{(R)}$
is given by  the Hartree term; i.e.
\begin{equation}
\varepsilon_{k}^{(R)} \rightarrow \varepsilon_{k}^{(R)} +
\frac{1}{N}\sum_{p} U^{(R)}_{k,p,p} n^{(FD)}_p.
\end{equation}
From the approximate solution (\ref{U_iter1}) we have
\begin{equation}
U^{(R)}_{k,p,p} = \frac{|v|^{2}}{\varepsilon_{k}+
\varepsilon_{p} - \Delta_{B}}
\end{equation}
which eventually leads to
\begin{eqnarray}
\varepsilon_{k}^{(R)} & \rightarrow & \varepsilon_{k}^{(R)} +
|v|^{2}\frac{1}{N} \sum_{p} \frac{n_{p}^{(FD)}}{\varepsilon_{k}+
\varepsilon_{p} - \Delta_{B}}
\nonumber \\
& = & \varepsilon_{k} + |v|^{2} \frac{1}{N}\sum_{p} \frac{n_{q=0}^{(BE)} +
n_{p}^{(FD)}}{\varepsilon_{k}+\varepsilon_{p} - \Delta_{B}}
\end{eqnarray}
and which thus is identical to the expression, Eqn. (\ref{ferm_perturb}) 
obtained from the  diagrammatic perturbation theory analysis.

\section{Numerical solution of the flow equations}

In this section we present the results of the selfconsistent
numerical solution of the flow equations. In order 
to solve the differential equations, Eqs. (\ref{hybrflow} - \ref{constflow}) 
we implement the Runge Kutta method.  The $l$ dependent
physical quantities are determined iteratively. 
Starting from their initial values, Eqs. 
(\ref{initial_conditions},\ref{constraints1},\ref{constraints2})
we determine them in the following step according to 
 $x(l+\delta l) = x(l)+\delta l x'(l)$, where $x'(l)$ stands 
for the derivative of
$x$ with respect to the flow parameter $l$, as given
by the corresponding flow equation. The increment which
we use for this procedure is the following: $\delta l=0.01$ for 
$l \leq 5$, $\delta l= 0.1$ for $5 < l \leq 100$, $\delta l = 1.0$ for $ 100 
< l \leq 1000$, and finally $\delta l= 10$ for $1000 < l \leq 10000$
(the flow parameter $l$ is expressed in units of the
inverse square bandwidth $D^{-2}$). The
major renormalizations take place up to $l \sim 100$ 
or $500$, but certain parts of the Brillouin zone are
slightly affected by  a further increase of $l$ up to
few thousands. We control our choice for the upper
limit of $l$ by: a) looking on the asymptotic behaviour
of all the renormalized quantities, and b) by checking
whether all the hybridization matrix elements $v_{k,q}(l)$ 
decreased below $0.1 \%$ of their initial value. 

 In order  to compare the results of this 
method  with the  results previously obtained by a selfconsistent 
perturbative treatment\cite{Ranninger-95}
we choose the same set of model parameters as used in the 
above mentioned previous studies, i.e., $v=0.1$,
$\Delta_{B}=-0.6$ and $n_{tot}=\sum_{\sigma} \left< c^{\dagger}
_{i\sigma}c_{i\sigma}\right> + 2 \left< b^{\dagger}_{i}
b_{i} \right> = 1 $ and consider  a one-dimensional tight
binding structure with the initial dispersion $\varepsilon_{k}
=-2t {\rm cos}(ka)$ (we set the lattice constant $a=1$ and
use the bandwidth $D=4t$ as a unit throughout this work).

\begin{figure}
\centerline{\epsfxsize=10cm \epsfbox{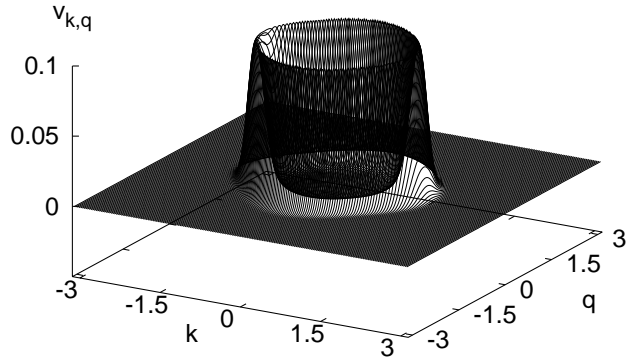}} 
\caption{Sketch of the hybridization matrix $v_{k,q}(l)$ for
$l=100$.}
\label{fig1}
\end{figure}

It is instructive to  first of all  have a look how the hybridization
matrix $v_{k,q}(l)$ evolves in the course of this renormalization technique.
In Fig. 1 we show $v_{k,q}(l)$ for $l=100$. Most of the
terms of the matrix (k,q) are practically reduced to zero while 
there is a region in the momentum space for which the
hybridization is only very weakly affected.

To understand why a situation like that shown in Fig. 1 takes place
let us come back to the approximate solution of the flow equations,
without treating the Fermion and Boson spectra  selfconsistently
(section II B). By inspection of Fig. (\ref{hybr_iter1})
we see that for $k$, $q$ such that $\varepsilon_{k} +
\varepsilon_{q} - \Delta_{B} = 0$ the coupling constant
$v_{k,q}(l)$ remains unrenormalized. In order to determine around which 
 momenta ($q$, $k$) this happens, we use $\varepsilon_{k}
\simeq -2t +tk^{2}$. This gives us a topology of concentric circles 
around a mean radius given by
\begin{equation}
 k^{*} \equiv \sqrt{k^{2} + q^{2}} 
\simeq 2 \sqrt{1+ \frac{\Delta_{B}}{D}}
= 0.894427 \;.
\label{patalogia}
\end{equation}
Solving the flow equations selfconsistently we do not encounter
such a pathological situation but still, near those $q$ and $k$ points, 
the renormalization of $v_{k,q}(l)$ evolves only very slowly.
There are different characteristic $l_{0}$ points from 
which efficient renormalization starts for the momentum sector near 
the resonant scattering $\varepsilon_{k}+\varepsilon_{q}=
E_{k+q}$ (as also pointed out before by Ragawitz and
Wegner \cite{Ragawitz-99} for the electron-phonon problem) and away from it.

Fig. 2 shows the evolution  of the hybridization constant
along the $q=-k$ cross-section as a function of l. It is clear
that renormalization of all the model parameters is necessary, 
otherwise the total elimination of the Boson-Fermion interaction 
is difficult or even impossible to fullfill.

\begin{figure}
\centerline{\epsfxsize=7cm \epsfbox{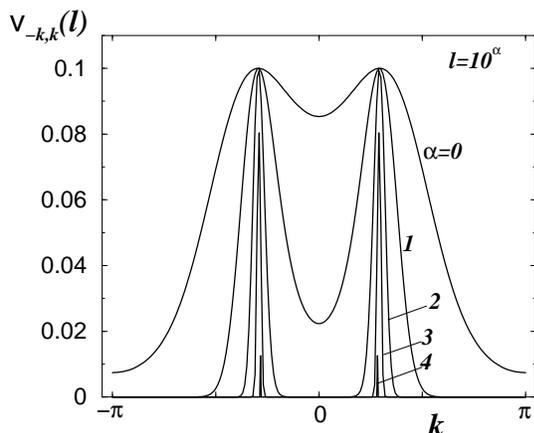}}
\caption{Evolution of the momentum dependence of the hybridization constant
$v_{-k,k}(l)$ as a function of the continuous  iteration parameter 
$l=10^{\alpha}$.}
\label{fig3}
\end{figure}

\vspace{0.5cm}
{\bf A. The evolution of the chemical potential}

\vspace{0.5cm}
To keep the total number of particles fixed we have to tune
the chemical potential. There are two effects observed in
the behavior of the chemical potential.
First, with a decrease of temperature, the chemical potential
approaches from below  the bottom of the Boson band.
Simultaneously, the bottom of the Boson band lowers and this
is the reason why below a given temperature ($T \sim 0.007$) 
$\mu$ starts to be pulled down. Of course, the relative distance 
$E_{0}-2\mu$ is a monotonously decreasing function of temperature.

In this $1-D$ case studied here, condensation of Bosons
can of course not take place. The chemical potential approaches 
asymptotically the lowest Boson level $E_{q=0}$  but never touches 
it except at $T=0$ (see Fig.3).

\begin{figure}
\centerline{\epsfxsize=7cm \epsfbox{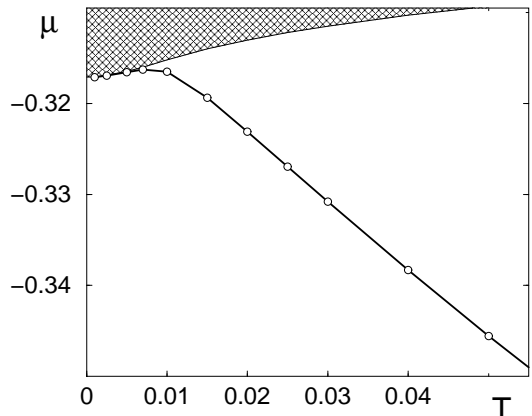}}
\caption{Variation of the chemical potential as function of
temperature (at $l=\infty$) for $\Delta_{B}=-0.6$, $n_{tot}=1$.
The Boson band is shown as a shaded area.}
\label{chempot}
\end{figure}

\vspace{0.5cm}
{\bf B. The Boson spectrum}

\vspace{0.5cm}
Due to the interaction with Fermions the initially localized
Bosons acquire itinerancy. The effective interaction between Bosons, 
being of order $v^4$, is hence neglected  here. In Figs. 4,5 we show
how, with a decrease of temperature, the Boson band becomes
broadened and the effective mass of the Bosons $m_{B}$
gets reduced, finally saturating around $m_{B}(T \rightarrow 0)
\sim m^{0}_{F}/4$. As a relative quantity we use $m^{0}_{F}
=\frac{\hbar^{2}}{2ta^{2}}$ which refers to the bare initial
Fermion mass for the $1 D$ tight binding case. A parabolic 
curvature of the long wavelength limit $k \rightarrow 0$ is 
characterized by the inverse effective mass.

For all temperatures studied by us the Boson dispersion
function exhibits a resonant like feature (a kink) around
momentum $q=2k^{*} \sim 1.8$. This does not correspond to the value 
of $2k_{F}$ as was initially mistakenly believed\cite{Ranninger-95}. 
By inspecting the
location of such a kink for other sets of model parameters we
conclude that it is mainly depending on the 
choice of $\Delta_{B}$. This kink occurs
for such momenta which satisfy the condition
\begin{equation}
\varepsilon_{k^{*}} + \varepsilon_{q-k^{*}} = E_{q} \;,
\end{equation}
where $q$ is a wavevector in  the first Brillouin zone.
Since within a precision of the order $v^{2}$ the value of the
Boson energies is around  the initial $\Delta_{B}$, $k^{*}$
is roughly determined by the condition $2\varepsilon_
{k^{*}}=\Delta_{B}$. In an approximate study of the flow
equations based on a  first iterative substitutions (section II. B)
one can check that the function $E_{q}$, Eq.(\ref{E_iter1})
in fact diverges  for $q=2k^{*}$.

\begin{figure}
\centerline{\epsfxsize=8cm \epsfbox{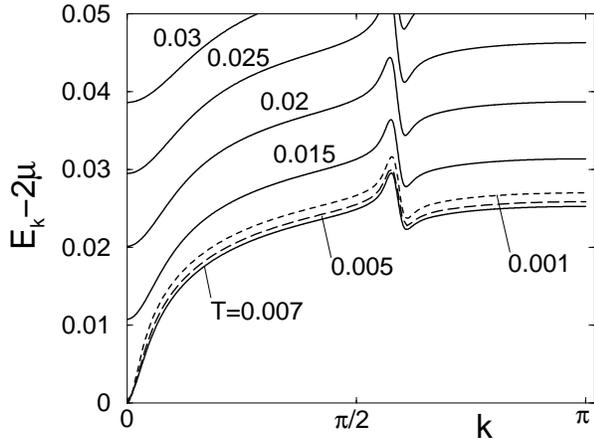}}
\caption{The effective Boson dispersion for a chosen set of
temperatures. Notice the increase of the  band width
with a decrease of temperature. At very small temperatures the
bottom of the Boson band asymptotically approaches the 
position of $2\mu$ and as a result the top of the Boson band
gets somewhat pushed up.}
\label{fig_Bosspec} 
\end{figure}

\begin{figure}
\centerline{\epsfxsize=8cm \epsfbox{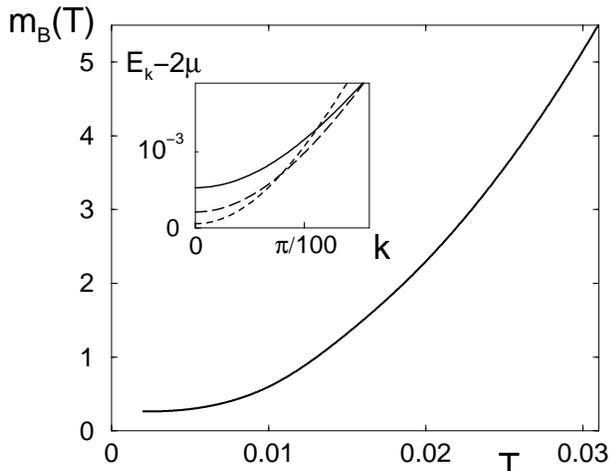}}
\caption{Low temperature behaviour of the effective Bosons
mass in units of the initial Fermions mass $m^{0}_{F}$. In the insert 
we draw the  long wavelength limit of the Boson  spectrum 
for the three lowest temperatures evaluated; $T= 0.001,  0.005 , 0.007$.}
\label{fig_mass}
\end{figure}

\vspace{0.5cm}
{\bf C. Interactions between Fermions}

\vspace{0.5cm}
As a result of the renormalization of the model parameters
we obtain an effective interaction between Fermions. Figs 6,7 
below illustrate the
momentum characteristics of the two channels defined in Eqs.
(\ref{BCS},\ref{density}). 

Again, we notice certain characteristic features appearing 
for the momenta 
corresponding to $k^{*}$. The interaction $U^{(d-d)}_{k,q}$
has a rather  regular behaviour: for  momenta such that
$k^{2}+q^{2} < k^{*2}$  this interaction is attractive, while
elsewhere it is repulsive. Around the region $k^{2}+q^{2}=
k^{*2}$ we observe a changeover, which looks quite singular.

The interaction $U^{(BCS)}_{k,q}$ shows  a similar behaviour
as $U^{(d-d)}_{k,q}$ but only along the cross-section $q=k$
(see the top of  Fig. \ref{U_cross}). Otherwise, the
corresponding change-over between the attractive and the repulsive
interaction regimes has a smooth character (see the bottom
of Fig. \ref{U_cross}). 

\begin{figure}
\centerline{\epsfxsize=10cm \epsfbox{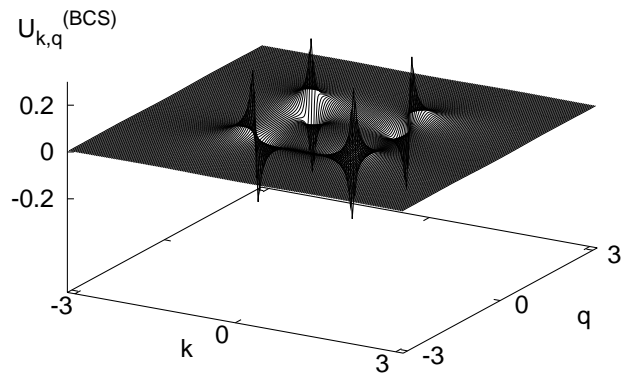}}
\caption{The BCS interaction strength $U^{(BCS)}_{k,q}$  for  $T=0.001$.}
\end{figure}
\begin{figure}
\centerline{\epsfxsize=10cm \epsfbox{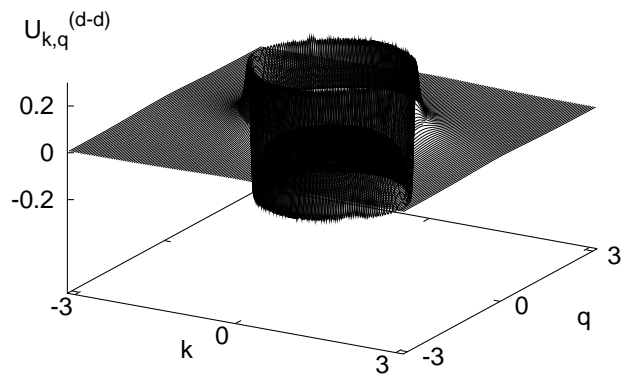}}
\caption{ the density-density interaction strength $U^{(d-d)}_{k,q}$ 
for $T=0.001$.}
\end{figure}

Temperature has a negligible
effect on the effective Fermion-Fermion interaction. For example,
 $U^{(BCS)}_{0,0}$ decreases only by about $0.4\%$ 
when the temperature is varied from $0.1$ to $0.001$. No qualitative
change is observed at all. Nevertheless temperature is an important 
factor in  as far as the  effectiveness
of this Fermion-Fermion interaction is concerned, since with 
a decrease of temperature $k_{F} \rightarrow k^{*}$.
To see that we refer the reader to consult Fig. \ref{chempot} 
keeping in mind that $E_{0} \simeq \Delta_{B}$. As seen from the 
Figs. 8 the interactions for 
$|k| < k^{*}$ have attractive character and are strongly 
enhanced (at least the elements $U_{k,q=k}$) infinitesimally
below the momentum $k^{*}$. One would naturally expect 
strong effects of these interactions if the Fermi vector
was located just below $k^{*}$.

\begin{figure}
\centerline{\epsfxsize=8cm \epsfbox{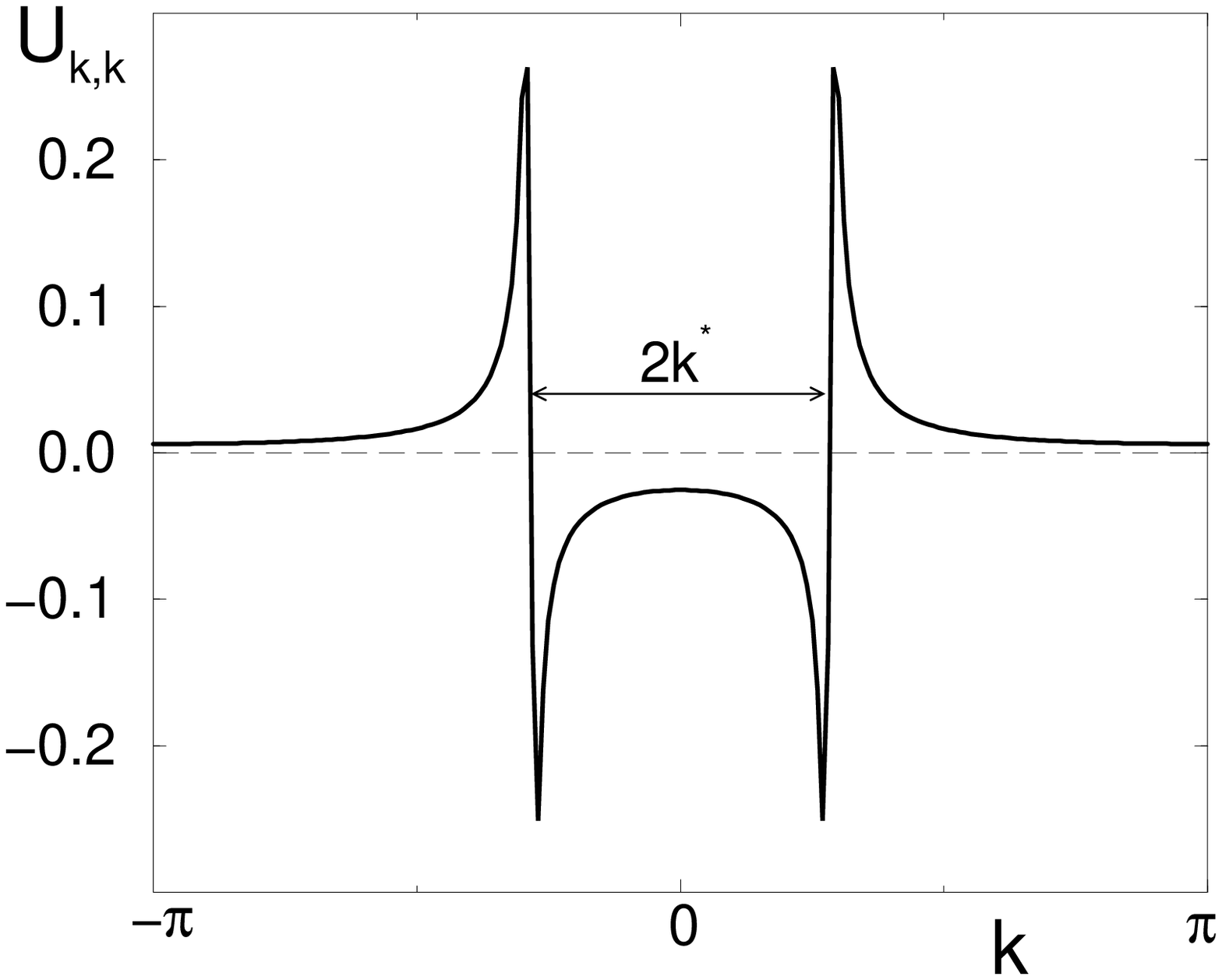}}
\centerline{\epsfxsize=8cm \epsfbox{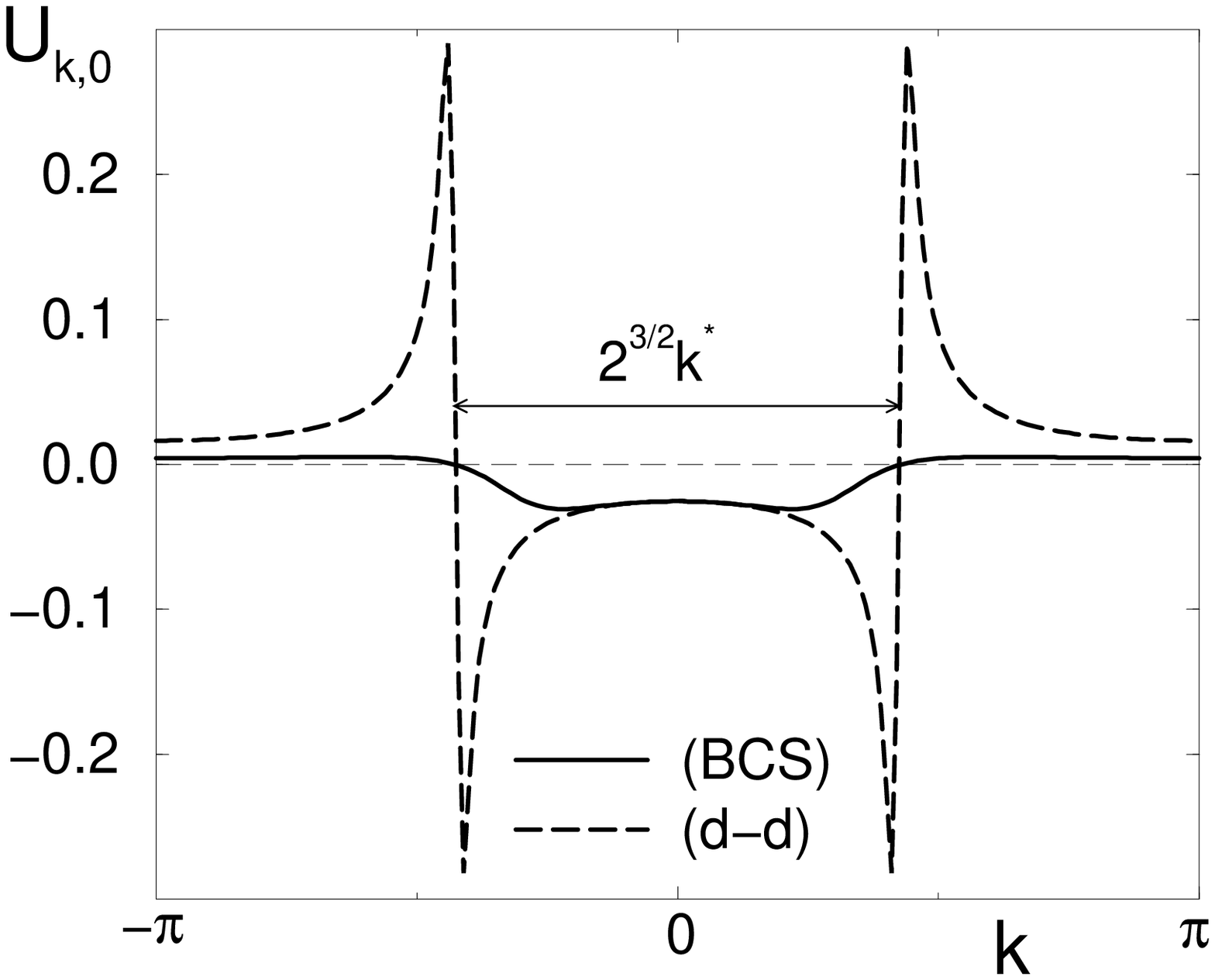}}
\caption{Cross-sections of the interactions $U_{k,q}^{(BCS)}$
and $U_{k,q}^{(d-d)}$ along a)  $q=k$ when both interactions
are identical (top figure), and b) for $q=0$ (bottom figure).
In general both potentials are repulsive for momenta $k >
k^{*}$ and become attractive if $k < k^{*}$.}
\label{U_cross}
\end{figure}

\vspace{0.5cm}

{\bf D. The Fermion spectrum}

\vspace{0.5cm}
Studying the Fermion spectrum is a rather complicated
issue because on one hand it is affected directly 
through the flow equations (17,18) and, on the other
hand, indirectly through the induced interactions as 
discussed  in the preceding section. The effect of temperature
turns out to be a very important factor in this problem.
In the physical regime of interest (i.e. when the effect of the
Boson - Fermion coupling manifests itself strongly) the
Fermi momentum is situated just  below $k^{*}$. 
Upon decreasing the temperature $k_F$ moves closer and closer 
to that value. As a result Fermion - Fermion interactions
become more and more effective and eventually may lead to
destruction of the quasi-particle nature  of those Fermions
in this model.

Below, we discuss what kind of fermion spectrum arises
purely on the basis of the renormalization of the free 
particle energies given in Eqs. (17,18). In Fig. 9 we plot 
the dispersion  $\varepsilon_{k}^{(R)}$ versus $k$
and in the inset its derivative with respect to initial 
$\varepsilon_{k}$. Using this derivative we compute the
density of states (DOS)
\begin{eqnarray}
\rho(\omega) &=& \frac{1}{N} \sum_{k} \delta \left( \omega + \mu
- \varepsilon^{(R)}_{k} \right)\nonumber \\ 
&=& \int d \varepsilon \rho^{0}(\varepsilon)
\left| \frac{d \varepsilon^{(R)}}{d \varepsilon} \right|^{-1}
\delta \left( \omega + \mu - \varepsilon^{(R)} \right)
\label{dos_def}
\end{eqnarray}
where $\rho^{0}(\varepsilon) = N^{-1}\sum_{k} \delta \left(
\varepsilon - \varepsilon_{k} \right)$ is the initial bare DOS of 
the Fermions.

\begin{figure}
\centerline{\epsfxsize=8cm \epsfbox{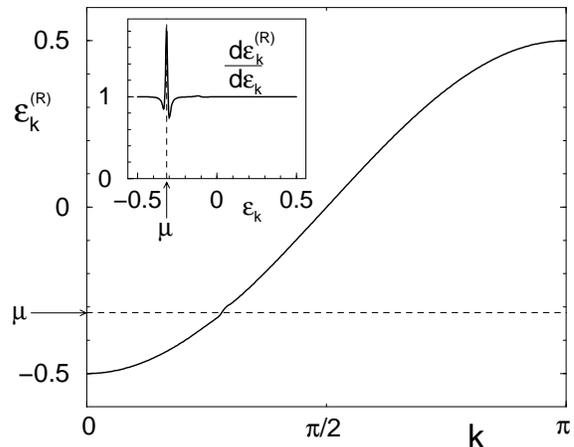}}
\caption{The renormalized ($l \rightarrow \infty$) Fermion
spectrum $\varepsilon^{(R)}_{k}$ for low temperature $T=0.005$.
Notice the tendency to form a gap around the Fermi vector.}
\end{figure}

We notice that below a characteristic temperature $T_{F}^{*}$
(at which the chemical potential starts to be pinned at the
bottom of the Boson band) there appears a pseudogap centered around
the Fermi level. In our case $T_{F}^{*} \sim 0.125$ in units
of the initial bare Fermion bandwidth.

An important question is the qualitative change-over of the 
pseudogap into a true BCS type gap below the condensation 
temperature $T_c$. This question can be tackled in a 
reasonably controlled way in the $1 D$ case. $T_c$ is then 
identically zero and at $T=0$ all Bosons are condensed in the $q=0$
state. Remember that we are dealing here with an effectively free 
Bose gas on a lattice, the Boson-Boson interaction - of order 
$v^4$ -  being neglected. 
We thus obtain
\begin{eqnarray}   
\frac{d\varepsilon_{k}(l)}{dl} & = & 4 n^{B}_{0}(l) 
\left( \varepsilon_{k}(l) - \mu(l) \right)
|v_{k,-k}(l)|^{2} \;,
\label{eps_T0}\\ 
\frac{dv_{k,-k}(l)}{dl} & = & - 4 \left( \varepsilon_{k}(l)
- \mu(l) \right)^{2} v_{k,-k}(l) \;,
\label{hybr_T0} \\
\frac{d \mu(l)}{dl} = \frac{1}{2} \; \frac{dE_{0}(l)}{dl} & = &
-2 \frac{1}{N} \sum_{k} | \varepsilon_{k}(l) - \mu(l) | \;
|v_{k,-k}|^{2}(l)
\label{mu_T0}
\end{eqnarray}
which follow from the general flow equations
(\ref{hybrflow}-\ref{Ekflow}) in the $T=0$ limit.
 $n^{B}_{0}(l)$ denotes the concentration of condensed 
Bosons and  $\mu(l)=E_{0}(l)/2$ in this limiting case. 
We need not find the whole Boson spectrum $E_{q}(l)$ 
because finite momentum states $E_{q\neq0}$ are not 
relevant in the ground state  (at least in absence of 
some external fields).

\begin{figure}
\centerline{\epsfxsize=8cm \epsfbox{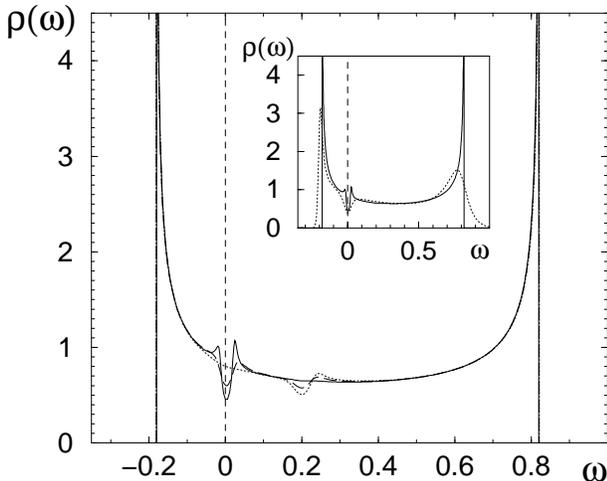}}
\caption{The DOS for the renormalized Fermion spectrum 
(\ref{dos_def}) for $T=0.02$ (dotted line), $T=0.01$ 
(dashed line) and $T=0.005$ (solid line). Upon lowering 
the temperature a pseudogap forms near the Fermi energy 
i.e., $(\omega=0)$, caused by the bonding two-Fermion 
states [22]. Some modification of the DOS 
is also seen for higher energies which corresponds to 
the antibonding two-Fermion state [22]. 
In the inset we compare the results obtained within 
the present flow equation technique (solid line) with 
those derived by selfconsistent diagrammatic techniques 
[5] (dotted line) for $T=0.005$.}
\label{dos}
\end{figure}

The set of Eqs.(\ref{eps_T0},\ref{hybr_T0},\ref{mu_T0})
is rather straightforward to study because the momentum dependence
of the hybridization constant $v_{k,-k}(l)$ enters only through
 $\varepsilon_{k}(l)$. Hence one can use the effective
DOS $\rho(\varepsilon,l)=1/N\sum_{k}
\delta(\varepsilon-\varepsilon_{k}(l))$ in Eq. (\ref{mu_T0}). 
We solve numerically these equations
for a fixed total concentration of particles which
is subject to the supplementary condition
\begin{eqnarray}
n^{B}_{0}(l) = \frac{1}{2} n_{tot} - \int_{-\infty}^{\mu(l)}
d\varepsilon(l) \; \rho(\varepsilon,l) .
\end{eqnarray}

\begin{figure}
\centerline{\epsfxsize=8cm \epsfbox{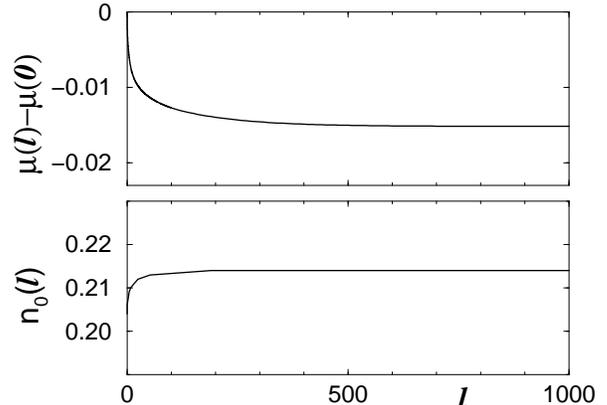}}
\caption{Evolution of the renormalized chemical potential
$\mu(l)$ (top) and concentration of Bosons $n^{B}_{0}(l)$
(bottom). Total concentration is kept fixed at $n_{tot}=1$.}
\label{fig11}
\end{figure}

\begin{figure}
\centerline{\epsfxsize=8cm \epsfbox{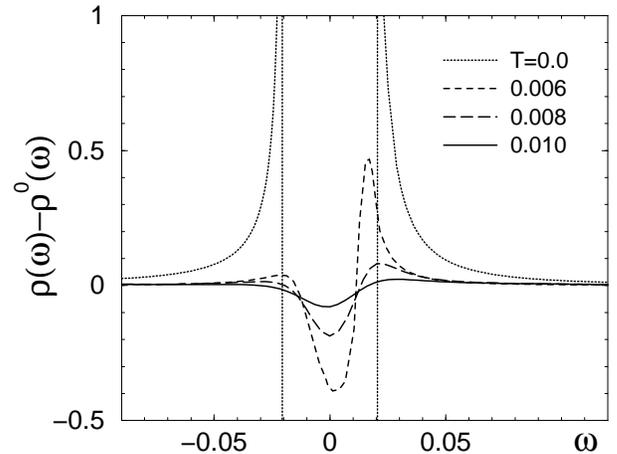}}
\caption{Difference between the DOS of the interacting 
$\rho(\omega)$ and of the free system $\rho^{0}(\omega)$ for 
$n_{tot}=0.6597$ and temperatures displayed in the legend.} 
\label{fig12}
\end{figure}

Fig. \ref{fig11} shows how the chemical potential is
renormalized through  this continuous transformation technique.
Notice that its variation is of the order $v^{2}$, as in
the case of finite temperatures. The total concentration of 
Bosons (which are all in the condensate state at $T=0$) is rather 
weakly affected by renormalization (of the order of 4 $\%$).

\begin{figure}
\centerline{\epsfxsize=8cm \epsfbox{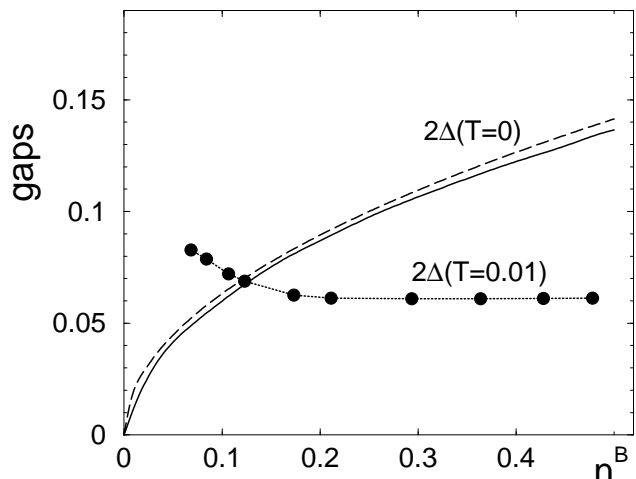}}
\caption{Variation of the true gap at $T=0$ as function
of Boson concentration obtained by the flow equation 
method (solid line) compared with the mean field solution
(dashed line). The filled circles show the corresponding
magnitude of the pseudogap obtained by the flow
equation method for a normal phase at $T=0.01$.} 
\label{fig13}
\end{figure}

The important outcome of this $T=0$ case analysis is seen
in the spectrum of Fermions. The asymptotical solution at
$l \rightarrow \infty$ yields a true gap in $\varepsilon_{k}^{(R)}$
which is formed around the chemical potential. The size of this
gap is in  very good agreement with the mean field
theory prediction, i.e. $\Delta(T=0)=v\sqrt{n^{B}_{0}}$ (see Fig. 
\ref{fig13}).

In Fig. \ref{fig12} we summarize our results obtained sofar in this 
section and which permit us to make some conjectures as to the 
evolution of the pseudogap into the true superconducting gap as 
the temperature is lowered.
First of all we notice two distinct energy scales which define these 
two gaps:

i) the superconducting gap being very sharp and being controlled by the 
first power in the coupling constant $v$

ii) the pseudogap, being evident in form of humps whose positions 
slightly  move closer together as the temperature decreases, 
varies with the second power of the coupling constant $v$ (remember 
that in the normal state our flow equation 
approach the results obtained are within the order of $v^2$). 

The relative size of the pseudogap and the zero temperature true gap 
can vary considerably as a function of the Boson concentration, as can 
be seen from Fig. \ref{fig13}. In Fig. \ref{fig12} we have chosen a 
situation close to the experimental situation in the high $T_c$ cuprates 
where the two gaps are of comparable size.

 In the 
next section we shall show that this same technique, applied to the 
superconducting state,  gives the superconducting gap varying with $v$.
It is clear that this difference in the size of these two gaps does not 
mean that they are of different origin! We shall come back to this question 
in the last section of this paper.

\section{The superconducting phase}

Bosons are not able to condense at any finite
(nonzero) temperature unless the dimensionality of the system
is higher than two. In this section we try to reach some preliminary
conclusions regarding the superconducting phase of the BFM 
 on the basis of flow equation method. We assume
that at least some fraction of the Bose subsystem is in the 
condensed state; in other words we consider $T<T_{BE}$.
We show  that the appearance of such a condensate
is inevitably related to the  formation of a true gap in
the Fermion subsystem. Our estimation of this gap
is in agreement with the mean field theory result for this
model \cite{Tomek}.

\vspace{0.5cm}
{\bf A. The Fermion Subsystem}

\vspace{5mm}
Given the existence of a certain fraction of condensed 
Bosons  we  put the chemical potential at the $q=0$ level
of the Boson energy spectrum, i.e., $\mu(l)=E_{q=0}(l)/2$. 
In the flow equation for Fermion energies, Eqs. (\ref{epsdownflow}) we
then have
 a dominating contribution coming from the $n^{(BE)}_{q=0}$
thermal factor and, as a consequence, can simplify this equation to
(which for $T=0$ is exact)
\begin{equation}
\frac{d\varepsilon_{k}(l)}{dl} \simeq  4 n^{B}_{0}(T,l)
\left( \varepsilon_{k}(l) - \mu(l) \right)
|v_{k,-k}(l)|^{2}.
\label{epsflow_super}
\end{equation}
 $n^{B}_{0}(T,l)$ denotes the concentration of condensed
Bosons at temperature $T$. This equation is coupled to the
flow equation for the hybridization coupling $v_{k,-k}$,
given in Eq.(\ref{hybr_T0}). By inspecting  the
bottom Fig. \ref{fig11}
we see that the $l$ dependence of the condensate concentration
can be dropped, $n^{B}_{0}(T,l) \simeq n^{B}_{0}(T)$.
For the strictly three-dimensional system the temperature evolution
is given through the standard relation  $n^{B}_{0}=
n^{B} \left[ 1 - \left(T/T_{BE}\right)^{3/2} \right]$,
where $n^{B}$ is the total  concentration of Bosons.
Similarly, the $l$ dependence of chemical potential can be
dropped because of the following arguments: 1) for
momenta close to the Fermi surface, the  renormalizations
of $\varepsilon_{k}(l)$ are of the order of $v\sqrt{n^{B}_{0}(T)}$
while the chemical potential undergoes a renormalizations of the
order $v^{2}$ (see  top Fig. \ref{fig11})) for  momenta which are 
far from $k_{F}$ (i.e. in
the high energy sector, using a terminology of the renormalization
group theory) renormalization is not effective at all
(except for the nevralgic $k^{*}$ point which is irrelvant
for the purpose of the present discussion).

Thus, without loss of  generality and precision,
we can rewrite  the flow equations in the
following from
\begin{eqnarray}
\frac{d \xi(l) }{dl} & = & 4n^{B}_{0} \xi(l) v^{2}(l) \\
\frac{d v(l) }{dl} & = & -4 \xi^{2}(l) v(l)  \;,
\label{constraint}
\end{eqnarray}
from which immediately follows 
\begin{equation}
n^{B}_{0}v^{2}(l)+\xi^{2}(l)= const.
\end{equation}
where $\xi(l)=\varepsilon(l)-\mu$  measures the Fermion energy
from the chemical potential.
It is evident from Eq. (50) that
for any nonzero $\xi$ the hybridization must evolve to
zero in the infinite $l$ limit, $v(\xi\neq0) \rightarrow 0$.
Consequently the renormalized spectrum becomes
\begin{equation}
\varepsilon^{(R)} -\mu = {\rm sign}\left( \varepsilon -
\mu \right) \; \sqrt{\left( \varepsilon -\mu \right)^{2}
+ n^{B}_{0}v^{2}} \;.
\label{constraint_eqn}
\end{equation}
The gap formed around the chemical potential in the Fermion subsystem
is given by
\begin{equation}
\Delta_{F}(T) = v \sqrt{ n^{B}_{0}(T)}
\end{equation}
which is the same as predicted by our previous studies \cite{PRB,Tomek}
of this model.  This result confirms that the BFM is characterized by a 
\underline{single} 
{\em transition temperature} $T_{c}$ at which the Bosons start
to condense and Fermions  to form a  gap in their excitation 
spectrum. Hence $T_{c}=T^{(B)}_{BE} = T^{(F)}_{BCS}$.

\vspace{5mm}
{\bf  B. The Boson Subsystem}

\vspace{5mm}
A very important issue in this context is to understand the  impact 
of the gap in the  Fermion spectrum on the excitation spectrum 
of the Bosons. Unfortunately the general flow equation, Eq.
(\ref{Ekflow}) can not be handled analytically,
not even in the  small $q$ limit.

In order to get some insight we study numerically
this equation together with the
constraint, Eq. (\ref{constraint}) which, evidently, applies only 
 for $T<T_{BE}$ and for dimensions larger than two. The set of 
equations we have then to solve are Eqs. (19,22) together with 
\begin{equation}
\left[ \varepsilon_{k}(l) - \mu(l) \right]^{2}  = 
\left[ \varepsilon_{k} - \mu \right]^{2} + n^{B}_{0}
\left[ v^{2} - | v_{k,-k}(l) |^{2} \right] \;
\label{epsk_eqn}
\end{equation}
which is a direct consequence of the constraint, Eq (51).
We fix the chemical potential at the level $\mu(l)=E_{q=0}(l)/2$ 
throughout the iterative solution procedure. 
Eq. (\ref{epsk_eqn}) controls  the formation of the gap in the Fermion
spectrum as a result of the presence of Boson in the condensate.
With the constraint, Eq. (\ref{epsk_eqn}) included, we 
investigate the Boson spectrum $E_{q}$ by restricting
ourselves to performing a one-dimensional momentum summation
in Eq. (\ref{Ekflow}). Such an approximative procedure 
reduces considerably the numerical complexity and is expected to 
give at least qualitatively correct results for dimensionality 
larger than two. A  complete numerical study on this point will be 
reported in some future work.

\begin{figure}
\centerline{\epsfxsize=8cm \epsfbox{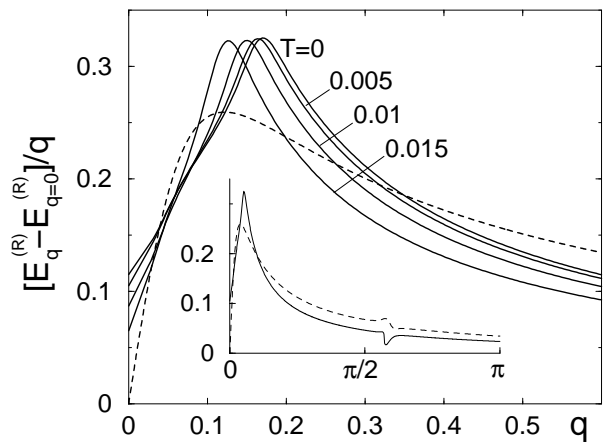}}
\caption{Momentum characteristics of the Boson excitation
spectrum in the superfluid state of the BFM with $n_{tot}=1$. 
Solid lines correspond to various
temperatures as marked and the dotted line is taken from
the selfconsistent solution for the normal phase at $T=0.005$.
Inset: the same function for the superfluid (solid line)
and normal phase (dotted line) within the half Brillouin
zone at $T=0.005$.}
\label{fig14}
\end{figure}

\begin{figure}
\centerline{\epsfxsize=8cm \epsfbox{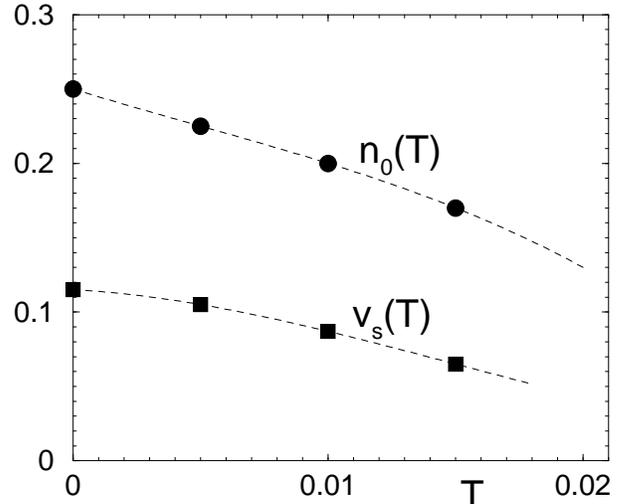}}
\caption{Sound velocity and the condensate concentration
versus $T$ for the superfluid phase of the BFM with $n_{tot}=1$.}
\label{fig_15}
\end{figure}

>From the numerical analysis of the above flow equations
we obtain  $E_{q}$. In the long wavelength limit  
($q \rightarrow 0$) it significantly deviates 
from its behavior, obtained above, in the normal state. 
In order to illustrate this, we plot in Fig. \ref{fig14} the 
momentum dependence of $\left[ E^{(R)}_{q}-E^{(R)}_{q=0}
\right] /q$ for several temperatures. Clearly the curves
show linear behavior up to momenta $q \sim 0.15$ and,
what is more important,  show a non-zero crossing point with
the ordinate. Its value determines the 
sound velocity $v_{s}(T)$ and marks the presence of a collective
excitation in the superfluid Bose subsystem. That such a sound
wave mode is completely absent in the normal phase can be seen by  
the dashed line in Fig. \ref{fig14} which crosses the ordinate at zero.

By inspection of Fig. \ref{fig14} one notices the decrease
of the sound velocity with increasing temperature.
Simultaneously the region of the $q^{2}$ behavior of the spectrum 
starts to shrink. In Figs. (15,16)
we show the dependence of the sound velocity
versus temperature (for a total concentration of carriers
$n_{tot}=1$) and versus total concentration at $T=0.0$.
These results agree well with the predictions for the
BFM obtained earlier by means of the dielectric
function formalism\cite{Tomek}. In the so-called
Bose limit, i.e. when the concentration of Bosons is not small,
the sound velocity has been shown to gradually decrease
with an increase of temperature towards $T_{c}$. On the other
hand, the sound velocity of the ground state gets reduced
when the total concentration of carriers increases (see Fig.
\ref{fig_15} of our present calculations and compare with 
Fig. 2 of  Ref.\ \cite{Tomek}). Only in the dilute regime
for small Boson  concentrations (the so called BCS limit)
one  expects a behavior, qualitatively different from that studied here.

\begin{figure}
\centerline{\epsfxsize=8cm \epsfbox{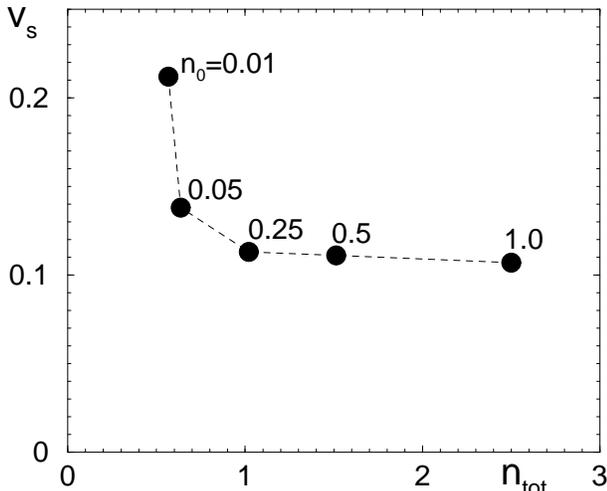}}
\caption{Sound velocity of the ground state $T=0$ as
a function of total concentration $n_{tot}$ for
$\Delta_{B}=-0.6$. Notice that critical concentration
above which superfluidity (superconductivity) can arise is
roughly equal $n_{cr} \sim 0.49$.}
\label{fig16}
\end{figure}

\section{Conclusion}
Our previous studies\cite{Ranninger-95} of the BFM indicated 
that the superconducting features of it arise  due 
to the initially localized Bosons becoming itinerant, an effect which is 
triggered by a
mutual feedback effect between the two subsystems. Concomitantely 
with this occurs the formation of a pseudogap in the DOS of the Fermions, 
as the temperature is lowered and which in turn permits the Bosons 
to acquire longer and longer lifetimes. The reason for that is 
a reduction of scattering processes due to a diminishing  number of 
Fermionic states available in this energy region. The opening of the 
pseudogap in those 
descriptions is linked to  a renormalized Fermion dispersion which
becomes flat as  the Fermi energy is approached from below, but, with at the 
same time, a substantial loss in spectral weight and lifetime 
broadening \cite{Ranninger-96}. The end result is in effect a separation 
into two separate subsystems, the Fermionic and the Bosonic one 
with their proper dynamics. Yet, due to the exchange coupling, they 
are mutually dependent on each other which leads to a single critical 
temperature, describing the onset of superconductivity, in both of the two
effective uncoupled subsystems.

The flow-equation technique studied in this work, renormalizes this 
inter-subsystem coupling to zero and hence is capable to make this 
interdependence of the dynamics of these two subsystems explicit. 
The various results obtained here are correct to second order in
the initial unrenormalized inter-subsystem coupling constant $v$.
The effective Bosonic subsystem behaves essentially as a lattice gas of free 
Bosons with a temperature dependent mass. The effective Fermion subsystem shows 
a dispersion which, upon approaching the Fermi wavevector 
rises almost vertically to some value above the Fermi energy within a 
very small regime around $k_F$. This reflects a pseudogap structure 
in the DOS which is of order $v^2$. In this regime of energies 
the effective renormalized intra-subsystem interaction is very singular 
(see Figs. (6 -\ref{U_cross})) and is expected to give rise to the lifetime 
effects and the reduction in spectral weight seen on our previous 
studies\cite{Ranninger-95}. For wavevectors greater than $k_F$ the 
effective Fermion dispersion remains quasi unrenormalized.

The use of the flow equation technique enables us to treat the 
superconducting and the normal state on the same footing and, 
on the basis of that, to make some conjectures as to how the pseudogap 
evolves into the true gap in the 
superconducting phase. For a $1D$ system where the superconducting phase 
is realized at $T \equiv 0$ the pseudogap evolves in from of a $V$-shaped 
curve which deepens until it touches the zero density level 
upon decreasing the temperature towards $T=0^+$. Upon entering the 
superconducting state at exactly $T=0$, this $V$ shaped curve changes 
abruptly into a  more conventional $U$ shaped curve, known from standard 
BCS type superconductors. The pseudogap is characterized by two distinct 
humplike features in the DOS whose positions get slowly closer to each other 
as  the  temperature is decreased.
The energy difference between those two humps being  of the order of $v^2$.
These pseudogap features are distinctively different from the gap structure 
in the superconducting state, as can be seen from Fig.12. The superconducting 
gap  shows a different variation with the inter-subsytem coupling 
constant, varying as $v$.
The  sizes for the pseudo- and the superconducting gap can in principle 
be quite different, as shown in Fig. \ref{fig13}. This does not  mean that 
they are of different physical origin.
We know from our previous studies\cite{Robin-98} that the pseudogap 
is very much independent on dimensionality, while the superconducting 
gap evidently is dependent on it. The present study, further, suggests 
that the variation of the two gaps with the concentration of Bosons varies 
in opposite direction. This can lead to a situation of a coexistence of 
two gap like structures in the superconducting phase, i.e., a 
superconducting gap and a remnant of the pseudogap. To what extent 
we should consider the pseudogap as a precursor of the superconducting 
is a question of semantics.
For the model system considered here, as well as for the real High $T_c$ 
cuprate materials the pseudogap is caused by amplitude rather than phase 
fluctuations.
Amplitude  fluctuations are a
prerequisite of the superconducting state but are not by themselves 
sufficient to guarantee its materialization. As a consequence, upon reducing 
the dimensionality of the system, the superconducting state is suppressed and 
tends to an insulating state with a characteristic upturn of the resistivity 
at low temperatures\cite{Robin-98}.

The study of the  change-over between the pseudogap and the superconducting 
gap for the more realistic anisotropic $3 D$ case, together with a careful 
study of the lifetime effects controlled by the intra-Fermion subsystem 
interactions is presently under investigation and  will be reported in 
some future study. 

Finally, for reasons of completeness, we should mention the 
theoretical studies 
of the pseudogap phenomenon based on the so-called BCS-Bose Einstein 
cross-over scenario. Such studies are based on effective BCS type coupling 
as well as the  negative U Hubbard model\cite{Hubbard}. Within this approach 
similar questions of a change-over form the pseudogap into a true gap have 
been considered\cite{PG-SG}. A discussion the precise differences between 
this scenario and the BFM scenario as concerns the physics and applicability 
to the High $T_c$ cuprates lies outside the frame of subject discussed in 
the present paper.

\section{Acknowledgement}
T.D. would like to acknowledge financial support from the
 University Joseph Fourier, Grenoble, and the hospitatily of the Centre 
de recherche pour les tres basses temperatures where this work 
was carried out. Moreover T.D. acknowledges  support from the Polish 
Committee of Scientific
Research under the grant No 2P03B10618.

\vspace{1cm}
{\Large \bf Appendix}

\vspace{5mm}
Let us consider the following modification of the generating
operator
\begin{equation}
\tilde{\eta} = \eta + \eta^{(2)}
\end{equation}
where $\eta$ is given by equation (\ref{generator})
and where we choose
\begin{equation}
\eta^{(2)} = \frac{1}{N} \sum_{\sigma} \sum_{p,k,q \neq k} 
\left( \gamma^{\sigma}_{p,k,q}(l) b^{\dagger}_{p+q} b_{p+k}
c^{\dagger}_{k \sigma} c_{q \sigma} - {\rm h.c.} \right).
\end{equation}
The coefficients $\gamma^{\sigma}_{p,k,q}(l)$ can be selected
in any arbitrary way provided that $v_{k,p}(l\rightarrow
\infty) \rightarrow 0$ still holds.

By a straightforward calculations one verifies that
\begin{eqnarray}
\left[ \eta^{(2)}, H \right] = \frac{1}{N} \sum_{\sigma} 
\sum_{p,k,q \neq k} \left( E_{p+k}-E_{p+q} -
\varepsilon^{\sigma}_{k}+\varepsilon^{\sigma}_{q}\right) 
\nonumber \\  \times 
\left( \gamma^{\sigma}_{p,k,q} b^{\dagger}_{p+q}
b_{p+k} c^{\dagger}_{k \sigma} c_{q \sigma}
+ {\rm h.c.} \right)
+O(\gamma v, \gamma U).
\end{eqnarray}
If we now use the Ansatz
\begin{eqnarray}
\gamma^{\downarrow}_{p,k,q}(l) & = & - \;
\frac{\varepsilon^{\downarrow}_{k}
+ \varepsilon^{\uparrow}_{p} - E_{p+k}}{E_{p+k}-E_{p+q} -
\varepsilon^{\downarrow}_{k} + \varepsilon^{\downarrow}_{q}}
\;\; v^{*}_{k,p}v_{q,p}
\label{e2a} \\
\gamma^{\uparrow}_{p,k,q}(l) & = & - \;
\frac{\varepsilon^{\downarrow}_{p}
+ \varepsilon^{\uparrow}_{k} - E_{p+k}}{E_{p+k}-E_{p+q} -
\varepsilon^{\uparrow}_{k} + \varepsilon^{\uparrow}_{q}}
\;\; v^{*}_{p,k}v_{p,q}
\label{e2b}
\end{eqnarray}
then we effectively obtain
\begin{eqnarray}
\left[ \eta^{(2)}, H \right] = -\frac{1}{N} \sum_{p,k,q\neq k}
b^{\dagger}_{p+q}b_{p+k} \left[ \left( \alpha_{k,p}+
\alpha_{q,p} \right) v^{*}_{k,p} v_{q,p}
\right. \nonumber \\ \times \left.
c^{\dagger}_{k\downarrow} c_{q\downarrow}
+ \left( \alpha_{p,k}+\alpha_{p,q} \right)
v^{*}_{p,k}v_{p,q} c^{\dagger}_{k\uparrow}c_{q\uparrow} \right]
+ O(v^{3}) \;.
\label{v3}
\end{eqnarray}
The terms containing $\gamma v$ become of the order $O(v^{3})$ 
and from the flow equation (\ref{interflow}) we can estimate 
$U \sim v^{2}$ which yields that $\gamma U \sim O(v^{4})$.

Thus on the right hand side of equation (\ref{v3}) we obtain the
same term as in (\ref{new_flow}) but with an opposite sign.
These terms subtract each other if one uses $\tilde{\eta}$
in the flow equation instead of the initial one (\ref{generator}).
The modified continuous unitary transformation does not generate
any interaction of the form $b^{\dagger}_{p+q}b_{p+k}
c^{\dagger}_{k\sigma}c_{q\sigma}$ for $q \neq k$ unless
hybridization constant $v$ is large enough.

\end{multicols}
\end{document}